\documentclass[a4paper,11pt]{article}
\usepackage{pos}

\title{Muon shower vertex reconstruction with waveform information in JUNO}

\author*[a]{Yongpeng Zhang}

\affiliation[a]{Institute of High Energy Physics, Chinese Academy of Sciences, 19B Yuquan Road, Beijing, China
\\On behalf of the JUNO Collaboration}

\emailAdd{ypzhang1991@ihep.ac.cn}

\abstract{The Jiangmen Underground Neutrino Observatory (JUNO) is a 20 kton liquid scintillator detector currently being built in a dedicated underground laboratory in China. It is a multi-purpose underground experiment with a physics program including neutrino mass hierarchy determination, precision measurement of neutrino oscillation parameters, measurement of solar, atmospheric, geo-neutrinos and other important neutrino physics searches. Electron anti-neutrinos are detected via the inverse beta decay by measuring the correlated positron and neutron signals. In this detection channel cosmic ray muon induced radioactive isotopes are the main background, especially those connected to cosmogenic backgrounds ($^{9}$Li/$^{8}$He and fast neutrons). They are predominantly produced by showing muons which account for about 10\% of all muons. Considering that the $^{9}$Li/$^{8}$He background is correlated with the parent muon in time and space, the vertex reconstruction of showers along the muon track is helpful to reject the backgrounds of $^{9}$Li/$^{8}$He and other isotopes. Based on the waveform simulation analysis, we know that the multi-peaks in waveform output by PMTs are mainly caused by these showers. Waveform analysis of muon events and preliminary results of shower vertex reconstruction based on detector simulation have been studied.}

\FullConference{%
  *** The 22nd International Workshop on Neutrinos from Accelerators (NuFact2021) ***\\
  *** 6–11 Sep 2021 ***\\
  *** Cagliari, Italy ***}

\begin{document}
\maketitle

\section{Motivation}

Most of rare events search experiments in the frontier of particle physics need to take a series of methods to reduce the background level and improve the sensitivity of physical targets. The neutrino detector used to study neutrino properties puts forward higher requirements for the background of the detector. Despite recent great progress in neutrino physics, there are certainly many open questions in neutrino physics and some intrinsic flavor issues of massive neutrinos will be addressed in the JUNO experiment.  For example, JUNO will accurately measure neutrino oscillation parameters and determine the neutrino  mass ordering (NMO) by detecting reactor anti-neutrinos. In addition, JUNO is also capable of exploring other physical yields.  However, in order to extract the mass ordering information from the spectral distortion of reactor neutrinos, the keys includes a large statistics with powerful sources (26.6 GW$_{th}$), an optimized distance to the reactor cores, an excellent energy resolution (3\%/$\sqrt{E}$), a good understanding of the energy response (energy non-linearity uncertainty of better than 1\%). Meanwhile, in order to improve the NMO sensitivity, it is necessary to lower the background as much as possible. According to the analysis results based on simulation so far\cite{PPNP}, almost half of the background comes from isotopes induced by cosmic rays.
\begin{figure}[h]
\centering
\includegraphics[width=0.4\linewidth]{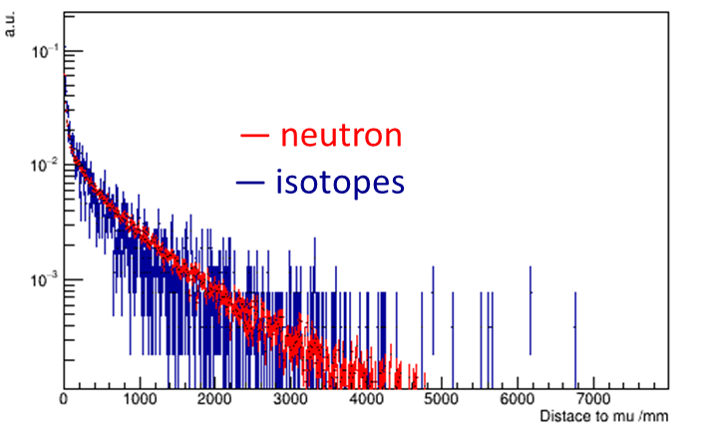}
\caption{\label{fig:distance}The distance distribution from each neutron and isotope to muon track. }
\end{figure}

Due to the strict requirements on background, larger overburden (~ 700 m) of rocks on top of the detector is needed to reduce the muon flux. Muon rate is estimated at about 0.004 Hz/m$^{2}$ and the average muon energy at about 207 GeV from simulation. In the liquid scintillator (LS), the energetic cosmic muons can interact with $^{12}$C and produce spallation neutrons and isotopes by electromagnetic or hadronic processes.  Muons that are accompanied by electromagnetic or hadronic showers, usually named as showering muons, are the dominant producers (>85\%) of the isotopes. Figure $\ref{fig:distance}$ shows the distance distribution from each neutron and isotope to muon track. Most of the isotopes and neutrons position are located within 3 m from the muon track.  The isotopes background can be effectively reduced by accurately reconstructing the muon track and veto a certain cylinder volume along the muon track.  With satisfactory isotope reduction efficiency, this veto strategy will introduce a relatively large dead volume.  Considering most isotopes produced by muon shower, the veto strategy which is making spherical volume veto at the vertex will be better if the shower vertex could be reconstructed. 

\section{Waveform analysis}

When a muon is going through LS, if a shower occurs, a large amount of energy will be deposited. Therefore, a large number of photons will be generated at the shower vertex and hit the PMTs. After waveform reconstruction, the distortion of waveform caused by the shower could be used to reconstruct the shower vertex.  Here, the waveform reconstruction is based on Toy MC by using the single photon electron waveform template which is formed when a photon hits the PMT.

\subsection{Waveform comparison of muon w/ and w/o shower }

Two kinds of events where generated: muons with a 10 GeV initial energy and muons with 200 GeV. All muons cross the detector at its center, with a vertical trajectory. Among the 10 GeV samples, the events with the smallest energy deposition are selected as the non-shower sample. We studied and compared the waveforms of PMTs coming from showering and straight muons, respectively, as shown in the left plot of fig.$\ref{fig:muon}$.  The curves in the fig.$\ref{fig:muon}$ are the waveform of non-shower muon and shower muon and are marked in red and black respectively. The blue curve is the difference between the waveform of non-shower muon and shower muon. The multi-peaks in waveform may be caused by the photons from shower, which need verification based Toy MC.

\subsection{Verification based on Toy MC }

\begin{figure}[h]
\centering
\includegraphics[width=0.4\linewidth]{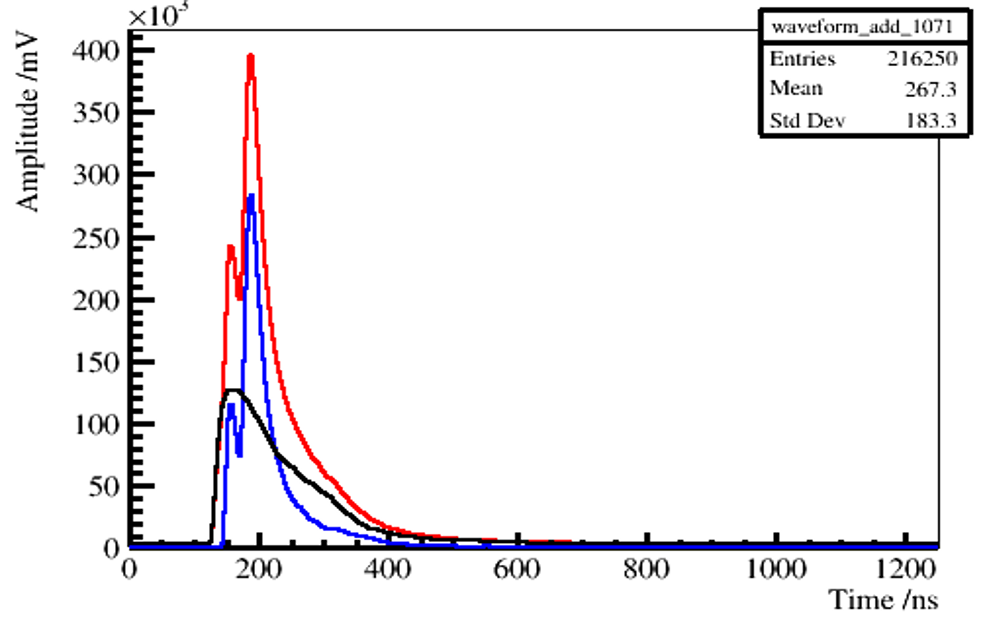}
\qquad
\includegraphics[width=0.4\linewidth]{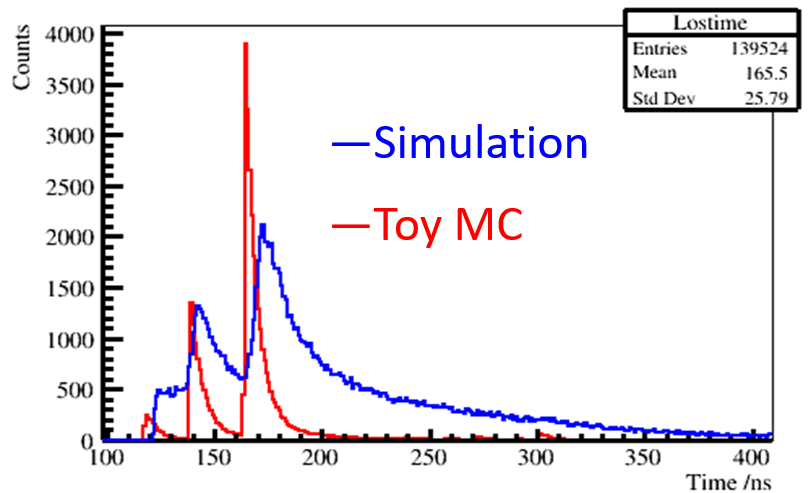}
\caption{\label{fig:muon} Left: The black and red curves are the waveforms of non-shower and shower muon, respectively. The blue curve is the difference between them. Right: The time distribution obtained from the time evolution based on Toy MC (red curve) and detector simulation (blue curve).
 }
\end{figure}

When muon have a shower in the LS, a large number of photons will be generated at the shower vertex. These photons propagate in the LS and then hit the PMTs. In the simulation, we can obtain the time and deposition energy of muon shower. The number of photons generated can be calculated according to the LS light yield. In addition, the number of photons hitting each PMT can be calculated considering the spatial azimuth probability of photons hitting PMT, LS attenuation length, quantum efficiency of PMT and other factors.  The photon hits time of the PMT is composed of the time when the shower occurs, the flight time of the photon propagating to the PMT and the LS luminescence time. The photon hit time distribution of the PMT is shown in right plot of fig.$\ref{fig:muon}$ and it was calculated by Toy MC based on the generation and propagation of photons.  The time distribution obtained from the time evolution based on the Toy MC is basically consistent with the hit time distribution of PMT in the simulation. Therefore, it can be concluded that the multi-peaks in waveform are indeed caused by the muon shower.
\section{Shower vertex reconstruction}
\subsection{Reconstruction process}

The multi-peaks information in waveform caused by the muon shower will be used to reconstruct the shower vertex. The number of peaks represents the number of shower vertices that can be reconstructed. With the predicted peaks time based on the time evolution from muon incident to photon hitting PMT and the observed peaks time in the waveform, the $\chi^{2}_{k_{th}}$ of k$_{th}$ peak in waveform can be built as:
\begin{equation}
\label{equ:chi}
\chi^{2}_{k_{th}} = \sum_{i}(\frac{T_{i}^{pre}-T_{i}^{Obs}}{\sigma_{i}})^{2}
\end{equation}

where  $T_{i}^{pre}$ is the predicted k$_{th}$ peak time, $T_{i}^{obs}$ is the observed (from data or MC) k$_{th}$ peak time and $\sigma_{i}$ represents the error of time for the $i^{th}$ PMT. The k$_{th}$ shower vertex parameters are obtained by minimizing the  $\chi^{2}_{k_{th}}$ function.

\begin{figure}[h]
\centering
\includegraphics[width=0.34\linewidth]{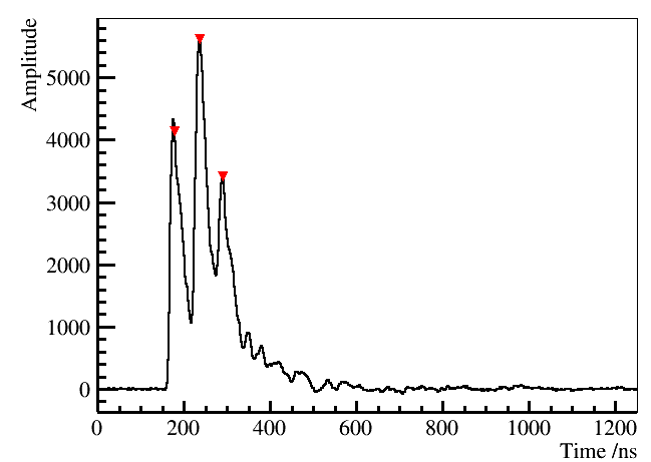}
\caption{\label{fig:waveform}3 larger peaks are marked by TSpectrum tool in the waveform of the single PMT. }
\end{figure}

The first step is to find the peaks of each PMT waveform. Considering that the single PMT has fluctuation at the tail of waveform mainly caused by photon reflection,  the waveforms of several locally adjacent PMTs are accumulated to smear the peaks of these fluctuations or only larger peaks are found. TSpectrum tool in the ROOT package is used to find the peaks in the waveform. For example, there are 3 larger peaks marked by TSpectrum tool in the waveform of the single PMT, as shown in fig. $\ref{fig:waveform}$. The second step is to calculate the predicted peak time with muon track information obtained by muon track reconstruction algorithm (such as the fastest light method).  The predicted peak time is composed of muon incident time, time of flight of muon, time of flight of photons and waveform rise time of the PMT.  The PMT which has the earliest photon hit time is regarded as the incident time.
Based on the predicted time and the observed peak time determined by peak seeking algorithm, the minimum $\chi^{2}$ values are calculated by iteration for each peak to obtain position parameters of each shower.

\subsection{Performance}

\begin{figure}[h]
\centering
\includegraphics[width=0.4\linewidth]{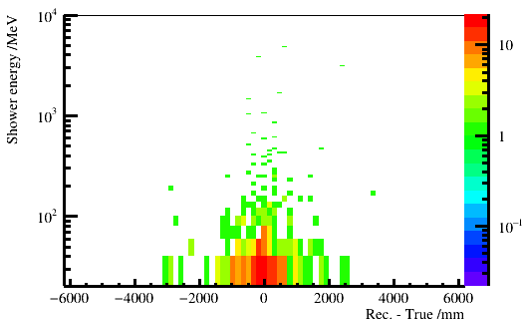}
\qquad
\includegraphics[width=0.35\linewidth]{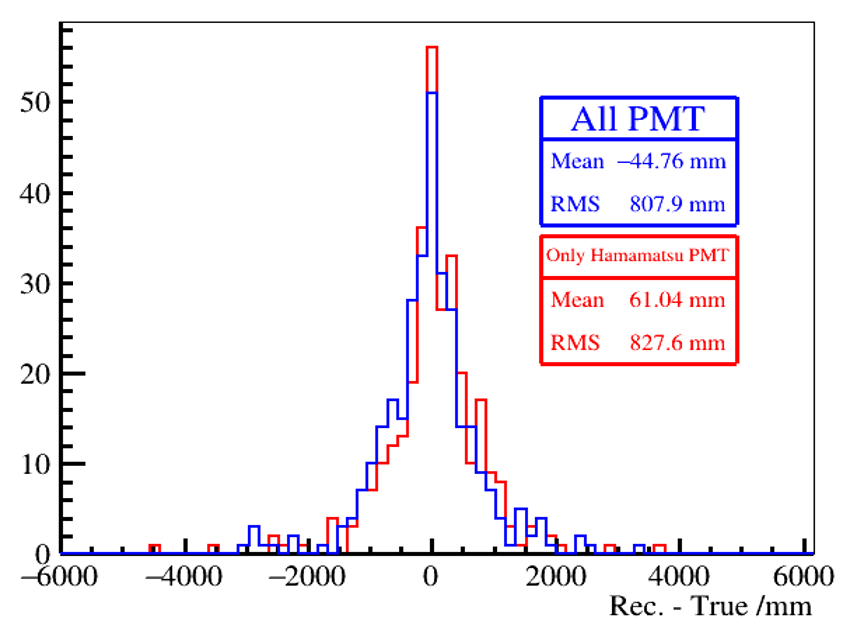}
\caption{\label{fig:result} Left:  The distribution between the shower energy and the position bias. Right:  The reconstructed results using different PMT type. }
\end{figure}

The preliminary reconstruction results are shown in the fig.$\ref{fig:result}$. The left figure in the fig.$\ref{fig:result}$ shows the distribution between the shower energy and the position difference between the reconstructed vertex and the true shower vertex.  The larger the shower energy, the smaller the position bias of the reconstruction vertex. The right figure in the fig.$\ref{fig:result}$ shows the difference between the reconstructed vertex and the true shower vertex obtained using different PMT type. Only using Hamamatsu PMTs has little impact on the results and shower vertex preliminary resolution is less than 1 m.

In order to evaluate the feasibility of reducing the isotopes background using the reconstructed shower vertex, a new shower muon veto strategy is proposed, which selects a spherical area of 3 m radius as veto around the shower vertex. Here, only a simple treatment is made to reduce the isotope background, that is, the veto strategy only considers the cut of spatial distance, and does not consider the selection cut of isotopes half-life. When calculating the veto efficiency, isotopes including $^{12}$B, $^{9}$Li, $^{8}$He, $^{9}$C, $^{8}$Li, $^{6}$He, $^{8}$B, $^{10}$C, $^{11}$C, $^{7}$Be etc. are considered. The preliminary veto isotopes efficiency is shown in table $\ref{tab:efficiency}$, where the reconstructed position resolution of shower vertex  with 1 m is considered in calculation. In addition, the neutron accompany efficiencies of cosmogenic isotopes are very higher based on the simulation.  Therefore, another veto strategy selects a 3 m sphere around each neutron position. As a comparison, the veto efficiency of the neutron veto method is also shown in the table $\ref{tab:efficiency}$, where the reconstructed position resolution of neutron with 10 cm is considered . It should be noted that the neutron accompany efficiencies of isotopes are model dependent.
\begin{table}[h]
\centering
\begin{tabular}{l c l c l}
\hline
\textbf{Method} & \textbf{Strategy} & \textbf{Veto efficiency}\\
\hline
Neutron veto & 3 m the spherical veto on neutron  & 99.7\%  \\ 
Shower vertex veto & 3 m the spherical veto on vertex & 94.8\% \\ 
\hline
\end{tabular}
\caption{\label{tab:efficiency} The veto isotopes efficiency and the veto strategy for shower vertex veto and neutron veto methods.   }
\end{table}

\section{Conclusion}

In the majority of underground rare-event experiments, spallation products constitute a severe background. Shower vertex reconstruction implemented based on the waveform contributes to the depression of isotopes background. The preliminary position resolution of vertex is less than 1 m and algorithms will be optimized to improve performance in the future. The larger shower energy, the better the reconstruction position resolution. The efficiency of shower vertex veto method which has no model dependency is slightly lower than that of the neutron veto method.

\end{document}